\documentstyle[osa,graphics,manuscript]{revtex}

\newlength{\figw}
\newlength{\figwl}
\newcommand{\mib}[1]{\mbox{\boldmath $#1$}}

\begin{document}    
\title{
  Quantum phase transition by cyclic four-spin exchange interaction for 
  $S=1/2$ two-leg spin ladder
}
\author{Yasushi Honda and Tsuyoshi Horiguchi}
\address{
  Dept. Computer and Mathematical Sciences, 
  Graduate School of Information Sciences, \\
  Tohoku University,
  Sendai 980-8579, Japan
}
\maketitle

\begin{abstract}
  We investigate an $S=1/2$ two-leg spin ladder with a cyclic four-spin 
exchange interaction whose interaction constant is denoted by $J_4$,
by using the density matrix renormalization 
group method.
The interchain and the intrachain interaction constant are denoted by 
$J_{\rm rung}$ and $J_{\rm leg}$, respectively and assumed to be 
antiferromagnetic.
It turns out that a spin gap between the singlet ($S_{tot}^z=0$) and the 
triplet ($S_{tot}^z=1$) states vanishes at $J_4/J_{\rm leg} \simeq 0.3$ for 
$J_{\rm rung}=J_{\rm leg}$.
This result is in contrast with the fact that the $S=1/2$ antiferromagnetic 
Heisenberg ladder,
that is the case of $J_{\rm rung} \neq J_{\rm leg}, J_4=0$, 
has a spin gap for all nonzero value of interchain interaction $J_{\rm rung}>0$.
We find a larger value of the correlation length 
for the spin-pair correlation function than a linear size $L$ of the system at 
$J_4/J_{\rm leg}=0.3$ and $J_{\rm rung}=J_{\rm leg}$:
the correlation length $\xi$ is about 204 times of the lattice constant for $L=84$
for these values of interactions.
We also find that the string correlation function decays rather algebraically 
than exponentially at $J_4/J_{\rm leg}=0.3$ and $J_{\rm rung}=J_{\rm leg}$.
These results suggest that there is a quantum phase transition at 
$J_4/J_{\rm leg} \simeq 0.3$ for $J_{\rm rung}=J_{\rm leg}$.
We estimate a phase boundary where the spin gap vanishes in a 
$J_4/J_{\rm leg}-J_{\rm rung}/J_{\rm leg}$ plane and obtain a consistent result with 
that by a perturbation theory for $J_{\rm rung}/J_{\rm leg}>1$.

\end{abstract}

\section{Introduction}
  In recent years, $S=1/2$ two-leg spin ladders have attracted 
considerable interests from both experimental and theoretical points
of view.
  The two-leg spin ladders are ideal models for quasi-one-dimensional 
materials such as SrCu$_2$O$_3$ \cite{Azuma94,Ishida94}, for which a spin gap has been 
observed.
Dagotto {\it et al.} \cite{Dagotto92} and Rice {\it et al.} \cite{Rice93} 
suggested moreover that hole doping to the two-leg spin ladders brings 
the superconductivity.
With respect to theoretical interest,
Barnes {\it et al.}\cite{Barnes93} 
suggested that the antiferromagnetic Heisenberg ladder with $S=1/2$ 
spins (AFHL) as shown in Fig.\ref{fig:afhl}
has a spin gap for all nonzero values of interchain interactions $J_{\rm rung}>0$.
The spin-pair correlation function of the AFHL decays exponentially, 
in contradiction to that of the $S=1/2$ antiferromagnetic Heisenberg chain,
which has a gapless spectrum and a power-law decay of the spin-pair 
correlation function.
It is believed that the resonating valence bond (RVB) picture is valid for
the ground state of the AFHL \cite{Rice93,Sigrist94,White94}.

We have another low-dimensional gapped system, that is the $S=1$ Heisenberg 
(Haldane) chain \cite{Haldane83}.
It is known as Haldane's conjecture that the antiferromagnetic Heisenberg
chains with integral spins are gaped, while the chains with half-integral 
spins are gapless \cite{Haldane83b}.
The valence bond solid (VBS) state of Affleck-Kennedy-Lieb-Tasaki (AKLT) model
\cite{Affleck87}
is believed to be an ideal example of the Haldane state of the $S=1$ system.
This exactly soluble AKLT model has biquadratic term 
$-\frac{1}{3}({\mib S_i}\cdot{\mib S_j})^2$ in addition to the usual Heisenberg 
Hamiltonian.
White \cite{White96} presented a numerical evidence for equivalence of the 
VBS (Haldane) state and the RVB state.
This implies that the spin gap in both the Haldane chain and the AFHL  
is due to the same mechanism.
Kolezhuk and Mikeska \cite{Kolezhuk97} also argued the equivalence of
the $S=1$ Haldane chain and an $S=1/2$ ladder which includes biquadratic terms
in addition to the usual bilinear terms.
They used a method of the matrix-product (MP) wave functions 
in order to discuss exact ground states of these systems and 
obtained phase transition points where
the spin gaps remain finite \cite{Kolezhuk98}.

For the $S=1/2$ two-leg spin ladder, it has been suggested that frustration brings 
a phase boundary where the spin gap vanishes \cite{Zheng98,Wang98}.
The frustration is due to bilinear terms which consist of the next-nearest-neighbor 
spins in the ladder.
It is known that a combination of bilinear and biquadratic terms 
can provide a cyclic four-spin exchange
interaction \cite{Honda93,Brehmer99}, whose interaction constant is denoted by $J_4$.
Brehmer {\it et al.}\cite{Brehmer99} pointed out that a moderate value of
$J_4$ for $J_{\rm rung}=J_{\rm leg}$ is consistent with experimental observations   
\cite{Eccleston98,Takigawa98,Magishi98,Imai98}.
Furthermore they found that the cyclic four-spin exchange interaction 
reduces an amount of the spin gap substantially.

The purpose of the present study is to clarify an effect of the cyclic four-spin
exchange interaction,
which consists of bilinear and biquadratic terms of spin operators and 
gives a frustration on the ladder, on the quantum phase transition.
We carry out the density matrix renormalization group
(DMRG) method for the $S=1/2$ two-leg spin ladder with the cyclic four-spin
exchange interaction and investigate the spin gap between the singlet 
and the triplet states.
We find that the spin gap vanishes at 
$J_4/J_{\rm leg} \simeq 0.3$ for $J_{\rm rung}=J_{\rm leg}$,
although we have finite spin gaps for $J_4/J_{\rm leg}<0.3$.
Furthermore 
at $J_4/J_{\rm leg} = 0.3$ for $J_{\rm rung}=J_{\rm leg}$ 
we observe 
a larger value of the correlation length for the spin-pair correlation function 
than the system size $L$ by assuming an exponential decay of the spin-pair 
correlation function.
We also find that the string correlation function decays rather algebraically
than exponentially at this point.
Those results suggest that there is a quantum phase transition at this point.
A phase boundary in the $J_4/J_{\rm leg}-J_{\rm rung}/J_{\rm leg}$ plane 
obtained in the present study is consistent with that obtained by 
a perturbation theory for $J_{\rm rung}/J_{\rm leg}>1$ \cite{Brehmer99}.

In section \ref{sec:model}, we express the $S=1/2$ two-leg spin ladder model
with the cyclic four-spin exchange interaction.
In section \ref{sec:results}, we present our results by the DMRG method:
we discuss the spin gap in subsection \ref{subsec:gap}, 
the spin-pair correlation function in subsection \ref{subsec:spin-pair},
the string correlation function in subsection \ref{subsec:string}
and the phase boundary in $J_4/J_{\rm leg}-J_{\rm rung}/J_{\rm leg}$ plane
in subsection \ref{subsec:phase_diagram}.
In section \ref{sec:conclusion},
we give conclusions of the present study.

\section{Two-leg spin ladder of $S=1/2$ with a cyclic four-spin exchange interaction}
\label{sec:model}
\begin{figure}
  \begin{center}
  \vspace*{10mm}
  \includegraphics{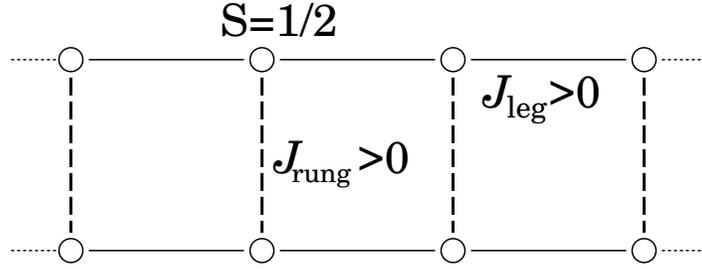}
  \end{center}
  \caption{\label{fig:afhl}Antiferromagnetic Heisenberg ladder with $S=1/2$ spin (AFHL).
     The Pauli spin operators are denoted by circles. 
     Antiferromagnetic interactions $J_{\rm leg}$ along each chain and $J_{\rm rung}$ 
     between the chains are denoted by solid and broken lines, respectively. 
            }
\end{figure}

\begin{figure}
  \begin{center}
  \includegraphics{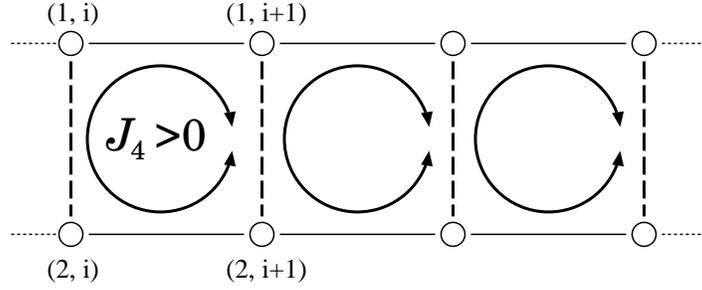}
  \end{center}
  \caption{\label{fig:afhl_ring}
     Antiferromagnetic Heisenberg ladder with a cyclic four-spin 
     exchange interaction $J_4$.}
\end{figure}

The antiferromagnetic Heisenberg ladder of $S=1/2$ (AFHL) is described by the 
following Hamiltonian:
\begin{eqnarray}
  {\cal H}_{\rm AFHL} = 
   && J_{\rm leg} \sum_i 
      ({\mib S}_{i,1} \cdot {\mib S}_{i+1,1} + {\mib S}_{i,2} \cdot {\mib S}_{i+1,2}) 
    + J_{\rm rung} \sum_i {\mib S}_{i,1} \cdot {\mib S}_{i,2} ,
  \label{eq:afhl}
\end{eqnarray}
where ${\mib S}_{i,1}$ and ${\mib S}_{i,2}$ express the Pauli spin operators on
chain 1 and 2, respectively.
Those operators are shown by circles in Fig.\ref{fig:afhl}.
The intrachain and the interchain interaction constant are denoted by $J_{\rm leg}$ 
and $J_{\rm rung}$, respectively, 
and are assumed to be antiferromagnetic: $J_{\rm leg}>0, J_{\rm rung}>0$.
They are shown by solid and broken lines in Fig.\ref{fig:afhl}, respectively.
We define a parameter $a$ by the ratio of these interactions as follows:
\begin{equation}
  a = J_{\rm rung}/J_{\rm leg} .
\end{equation}
The Hamiltonian with the cyclic four-spin exchange interaction is described
as follows:
\begin{equation}
  {\cal H}= {\cal H}_{\rm AFHL}+J_4 \sum_i (P_{4,i}+P_{4,i}^{-1}),
\end{equation}
where $P_{4,i}$ means a permutation operator of four spins.
This operator expresses a cyclic permutation of four spins clockwise:
$(1,i) \rightarrow (1,i+1) \rightarrow 
(2,i+1) \rightarrow (2,i) \rightarrow (1,i)$ (see Fig.\ref{fig:afhl_ring}).
The inverse operator $P^{-1}_{4,i}$ expresses the permutation of those four spins 
counterclockwise.
The sum of these permutation operators is expressed by a sum of spin operators 
as follows:
\begin{eqnarray}
  P_{4,i}&&+P_{4,i}^{-1}
    = \nonumber \\
   && 4({\mib S}_{i,1}\cdot{\mib S}_{i+1,1})
     ({\mib S}_{i,2}\cdot{\mib S}_{i+1,2}) +4({\mib S}_{i,1}\cdot{\mib S}_{i,2})
     ({\mib S}_{i+1,1}\cdot{\mib S}_{i+1,2}) 
    -4({\mib S}_{i,1}\cdot{\mib S}_{i+1,2}) ({\mib S}_{i+1,1}\cdot{\mib S}_{i,2}) 
    \nonumber \\
    &&+( {\mib S}_{i,1}\cdot{\mib S}_{i+1,1}
      +{\mib S}_{i,2}\cdot{\mib S}_{i+1,2} +{\mib S}_{i,1}\cdot{\mib S}_{i,2}
      +{\mib S}_{i+1,1}\cdot{\mib S}_{i+1,2}) 
             \nonumber \\
    &&+( {\mib S}_{i,1}\cdot{\mib S}_{i+1,2}
      +{\mib S}_{i+1,1}\cdot{\mib S}_{i,2} 
     )\nonumber \\
   &&+1/16 .
\end{eqnarray}
In this way, the sum of the permutation operators is expressed by a combination of 
bilinear terms and biquadratic terms of intrachain spins and interchain spins.

\section{Results by the DMRG method}
\label{sec:results}
In the present study, we use the infinite system algorithm of the DMRG method.
All of the data calculated in the present study are obtained by using the open
boundary conditions.
By dividing the present section into 4 subsections, we present our results
as for the spin gap, the correlation length of spin-pair correlation function,
the string correlation function and the phase diagram in turn.
\subsection{The spin gap}
\label{subsec:gap}
\begin{figure}
\begin{center}
  \includegraphics{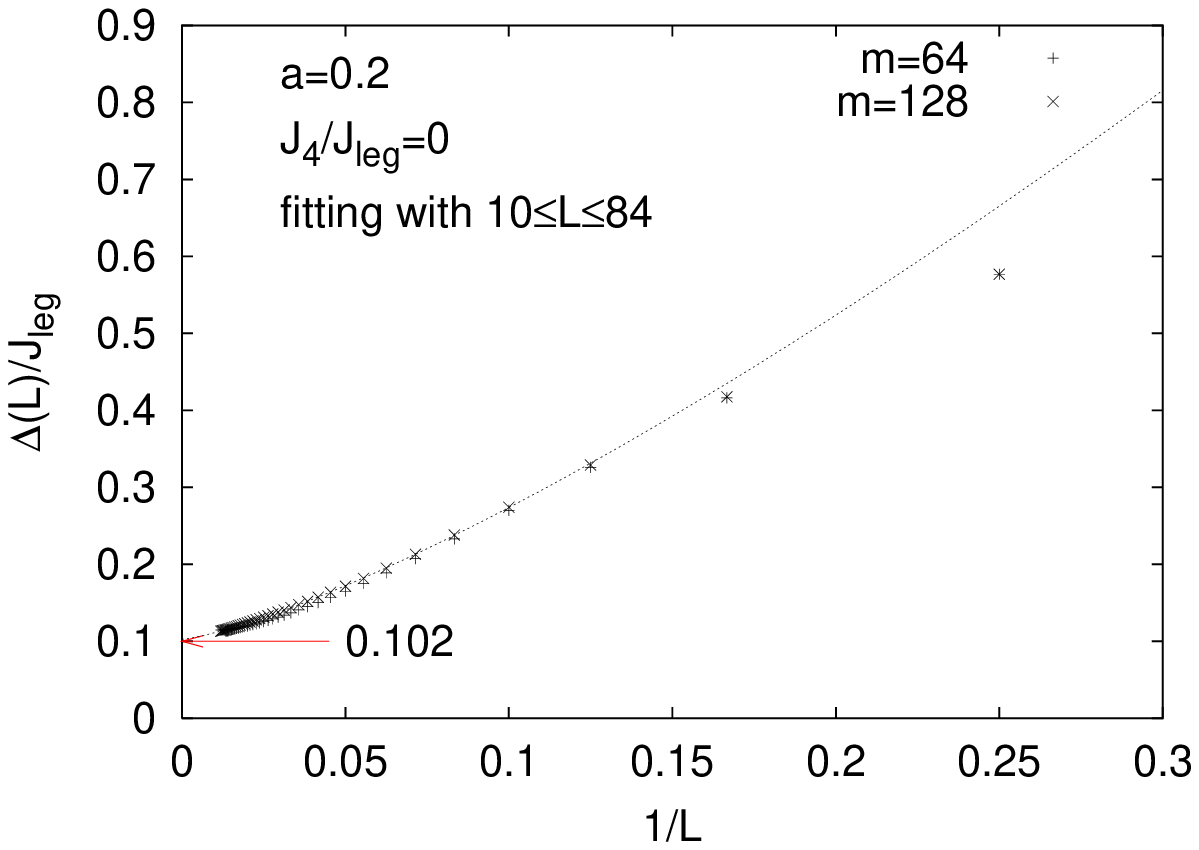}
\end{center}
\caption{\label{fig:gap_a5J400_Ldep}
   $\Delta(L)/J_{\rm leg}$ as a function of $1/L$ 
   for $J_4/J_{\rm leg}=0$ and $a=0.2$.}
\end{figure}

We define the spin gap as follows:
\begin{equation}
  \Delta(L) = E_0(L,1)-E_0(L,0),
\end{equation}
where $E_0(L,0)$ and $E_0(L,1)$ are the lowest energies for which the total
value of $z$-component of the spin operator, namely $S^z_{\rm tot}$, is 0 or 1,
respectively, for the ladder of length $L$;
there are $2L$ sites in the ladder.
We show size dependences of $\Delta(L)/J_{\rm leg}$
for $J_4/J_{\rm leg}=0$ and $a=0.2$ as a function of $1/L$
in Fig.\ref{fig:gap_a5J400_Ldep}.
%
The values of $m$ in Fig.\ref{fig:gap_a5J400_Ldep} show the number of eigenstates of 
the density matrix which are kept in the DMRG method.
Difference between the values of the spin gap calculated for $m=64$ and 128
is within the size of symbols plotted in Fig.\ref{fig:gap_a5J400_Ldep}.
We estimate the value of the spin gap in the thermodynamic limit $L \rightarrow \infty$
using the values for $m=128$.
In this extrapolation,
the values for $10\leq L \leq 84$ are used in order to avoid the effect of small 
system size;
we obtain the spin gap $\Delta(\infty)/J_{\rm leg}\simeq 0.102$.

\begin{figure}
\begin{center}
  \includegraphics{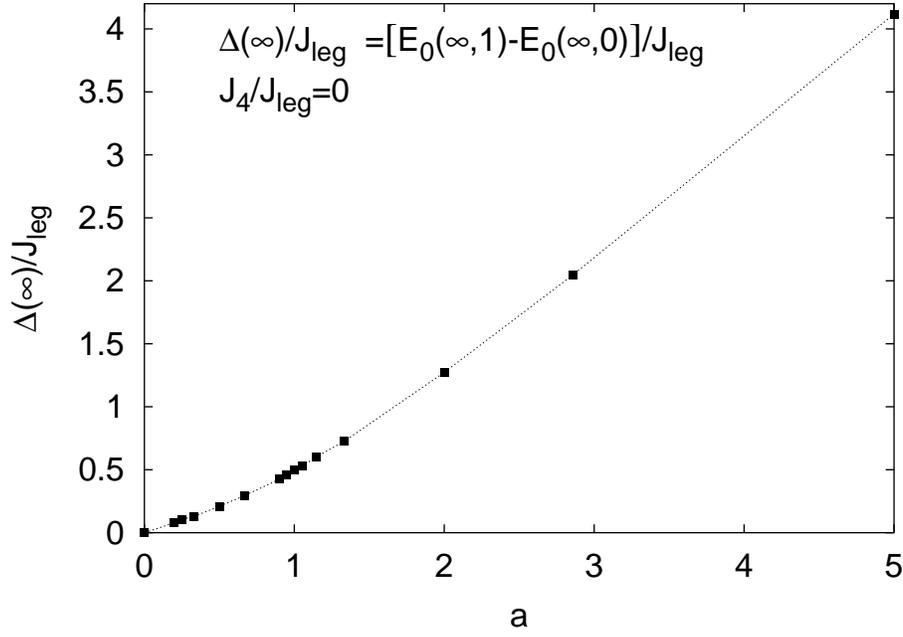}
\end{center}
\caption{\label{fig:gap-inval}
   Extrapolated value of spin gap for $J_4/J_{\rm leg}=0$ 
   as a function of $a$.}
\end{figure}

\begin{figure}
  \begin{center}
  \includegraphics{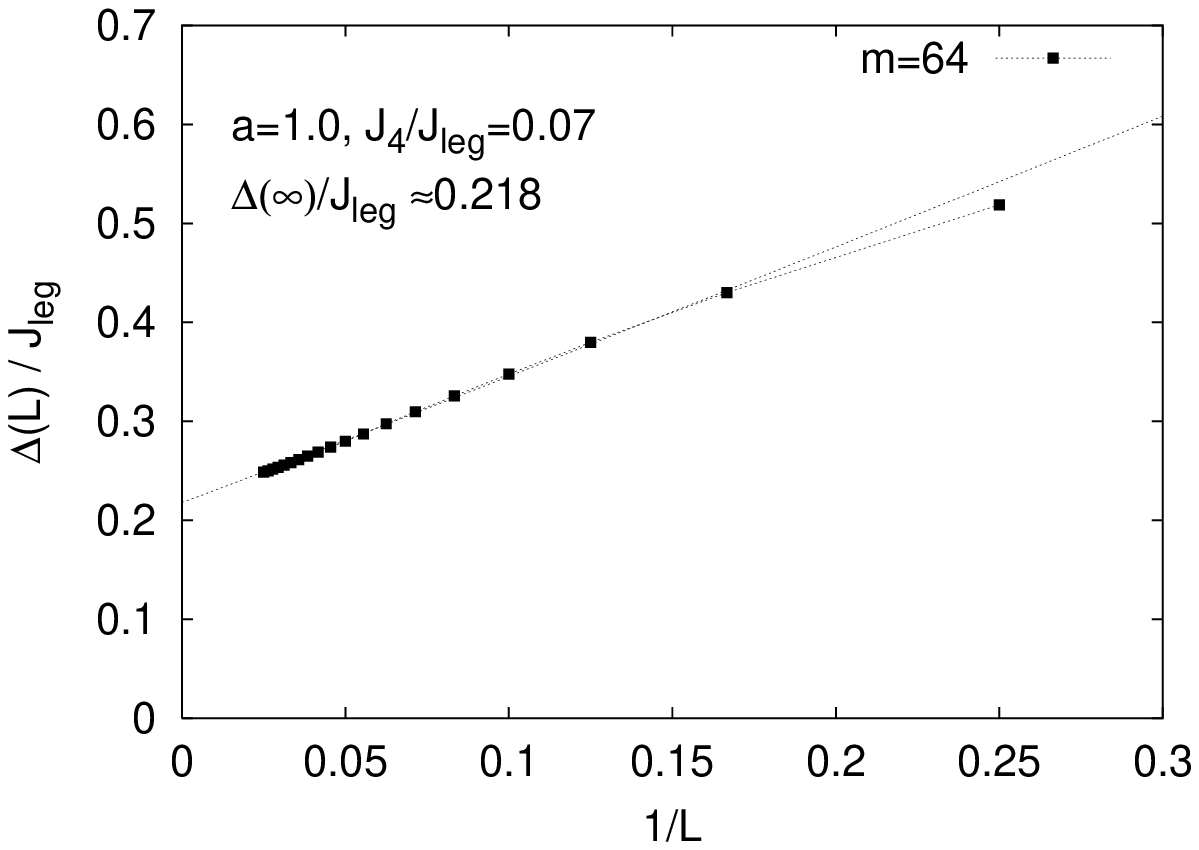}
  \caption{\label{fig:gap_a1J4007_Ldep}
     Size dependence of the spin gap for $J_4/J_{\rm leg}=0.07$ and $a=1$.
  }
  \end{center}
\end{figure}

In Fig.\ref{fig:gap-inval}, we show the extrapolated values of the spin gap 
$\Delta(\infty)/J_{\rm leg}$ for $J_4/J_{\rm leg}=0$
as a function of $a$.
The case of $a=1$ and $J_4/J_{\rm leg}=0$ corresponds to the AFHL with 
$J_{\rm rung}=J_{\rm leg}$.
On the other hand, the case of $a=0$ corresponds to two independent
chains since there is no interchain interaction.
We do not find a gapless region for $a>0$.
The present result by the DMRG method is consistent with the assertion 
given by Barnes {\it et al.} \cite{Barnes93}.
In the standard analysis of experimental data for neutron scattering,
nuclear magnetic resonance(NMR) and nuclear quadrupole resonance (NQR)
\cite{Eccleston98,Takigawa98,Magishi98,Imai98},
$a \sim 0.5$ is expected by assuming $J_4=0$.
We obtain the spin gap $\Delta(\infty) / J_{\rm leg} \simeq 0.21$
for $J_4/J_{\rm leg}=0$ and $a=0.5$.
%
%
On the other hand, $J_4/J_{\rm leg}$ is estimated to be 0.07 from comparison 
between experimental data and numerical results given in Ref.\ref{ref:Brehmer99}.
In Fig.\ref{fig:gap_a1J4007_Ldep}, the spin gap 
for $J_4/J_{\rm leg}=0.07$ and $a=1$ is shown as a function of $1/L$.
The behavior in Fig.\ref{fig:gap_a1J4007_Ldep} is a typical example of size 
dependence of the spin gap in the DMRG method,
which occurs for the system with a finite 
value of the spin gap.
We estimate the spin gap for $J_4/J_{\rm leg}=0.07$ and $a=1$ to be 0.218
in the thermodynamic limit.
We notice that the both cases, $a=1, J_4/J_{\rm leg}=0.07$ 
and $a=0.5, J_4/J_{\rm leg}=0$ give similar values for the spin gap.
Hence there is a possibility that 
the assumption $J_4=0$ is not valid for the analysis of results for NMR or NQR.

\begin{figure}
  \begin{center}
  \includegraphics{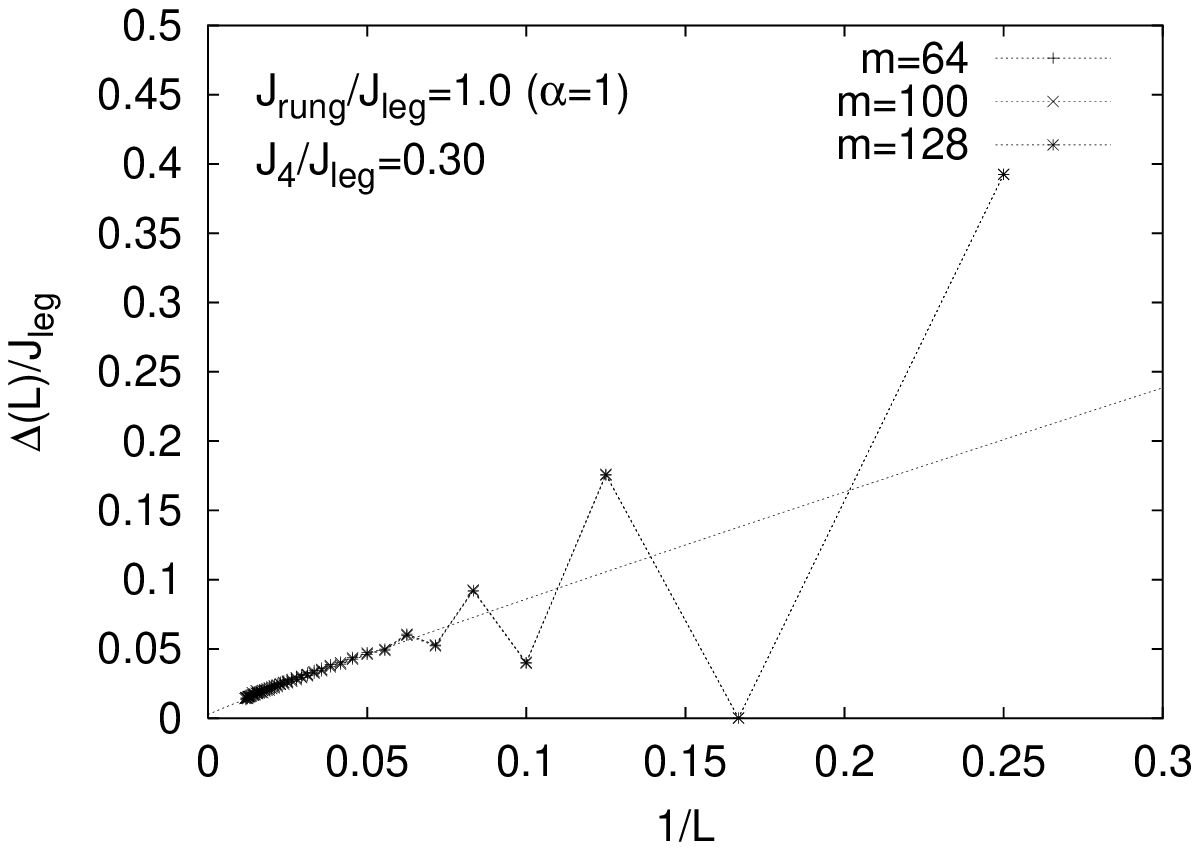}
  \end{center}
  \caption{\label{fig:gap_a1J403_Ldep}
    Size dependence of the spin gap for  
    $J_4/J_{\rm leg}=0.3$ and $a=1$.}
\end{figure}

In our calculations, the spin gap decreases as the value of $J_4$ increases.
We show the result for $J_4/J_{\rm leg}=0.3$ and $a=1$
in Fig.\ref{fig:gap_a1J403_Ldep}.
%
We find that the values of the spin gap for $L=4n$ and for $L=4n+2 \ \ (n=1,2,\cdots)$ 
show different behavior as for size dependence, especially for smaller
system sizes.
Therefore, the data for $L < 20$ are discarded for an
extrapolation to $L \rightarrow \infty$.
Hence we obtain $\Delta(\infty)/J_{\rm leg}=0.00 \pm 0.003$ for 
$J_4/J_{\rm leg}=0.3$ and $a=1$.

\begin{figure}
  \begin{center}
  \includegraphics{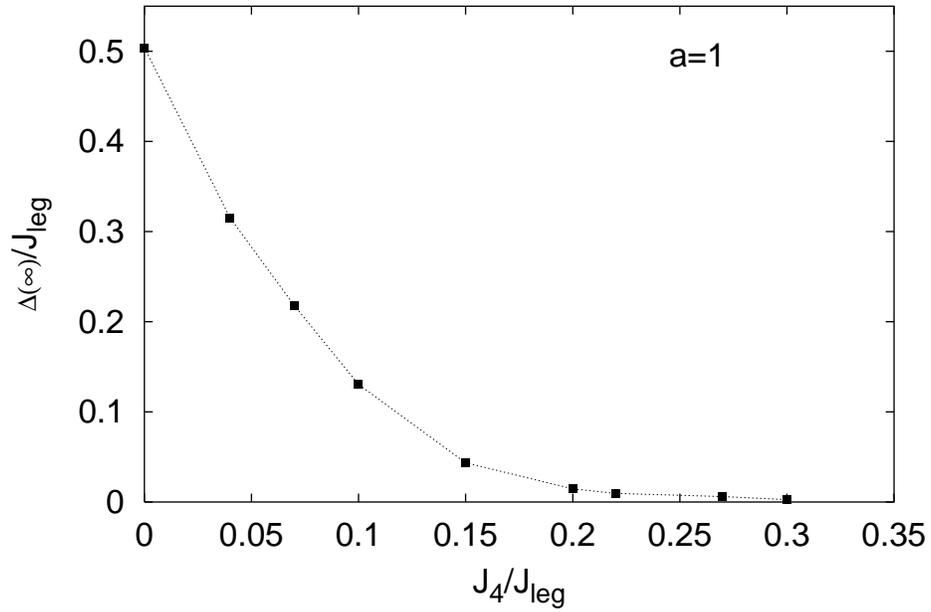}
  \end{center}
  \vspace*{-7mm}
  \caption{\label{fig:gap_a1J4dep}
    $J_4/J_{\rm leg}$ dependence of the spin gap $\Delta(\infty)/J_{\rm leg}$ for $a=1$.}
\end{figure}

We show the value of spin gap as a function of $J_4/J_{\rm leg}$ 
for $a=1$ in Fig.\ref{fig:gap_a1J4dep}.
As the value of $J_4/J_{\rm leg}$ increases, the value of spin gap decreases gradually.
We notice that the spin gap has very small values
in the region $J_4/J_{\rm leg} \gtrsim 0.2$ and becomes zero near 
$J_4/J_{\rm leg} \sim 0.3$.

\subsection{The Correlation length of the spin-pair correlation function}
\label{subsec:spin-pair}
\begin{figure}
 \vspace*{5mm}
 \begin{center}
 \includegraphics{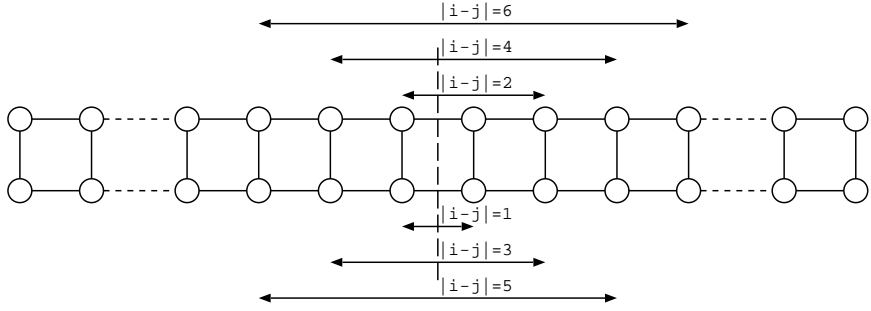}
 \vspace*{5mm}
 \caption{\label{fig:corr}
    A choice of sites $i$ and $j$. We assume that they are located symmetrically
    with respect to the center of the ladder.  
 }
 \end{center}
\end{figure}

\begin{figure}
  \begin{center}
  \includegraphics{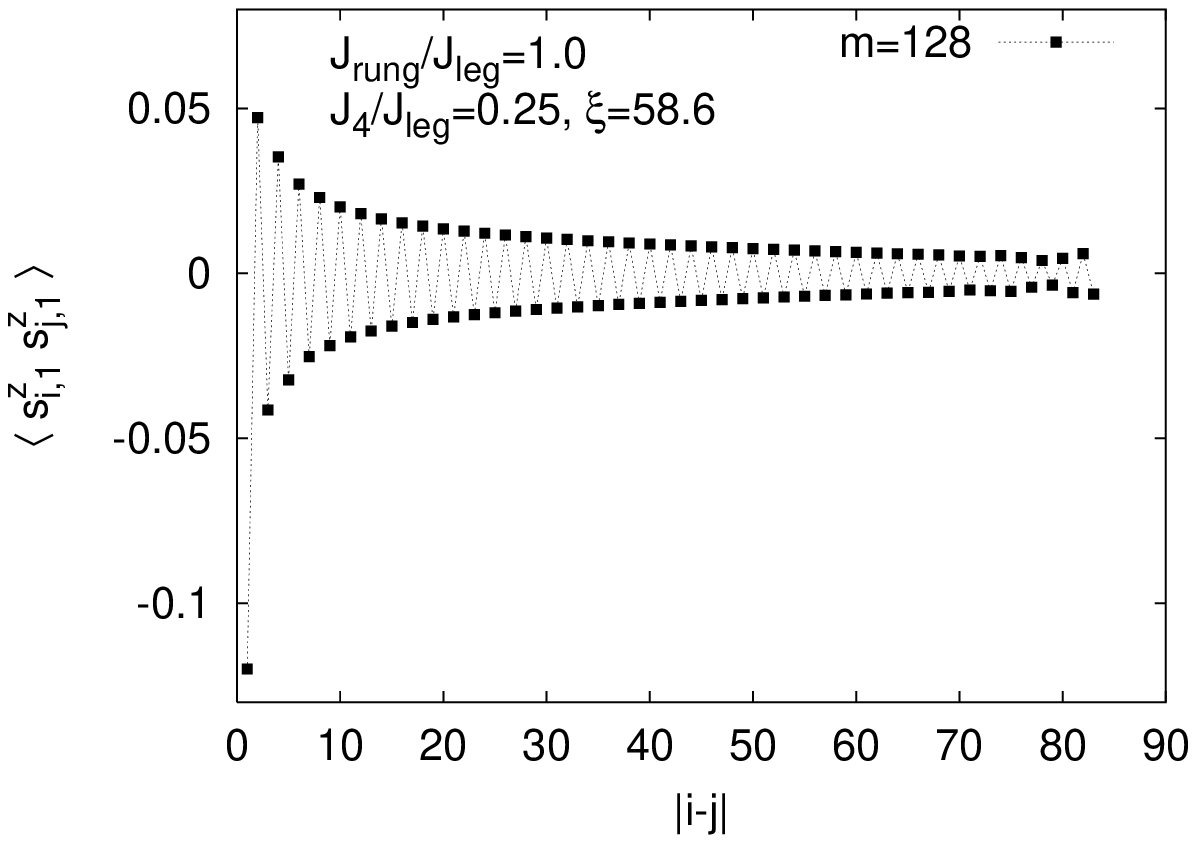}
  \vspace*{-0.38\figwl}\\
  \hspace*{0.38\figwl}
  \scalebox{0.4}{\includegraphics{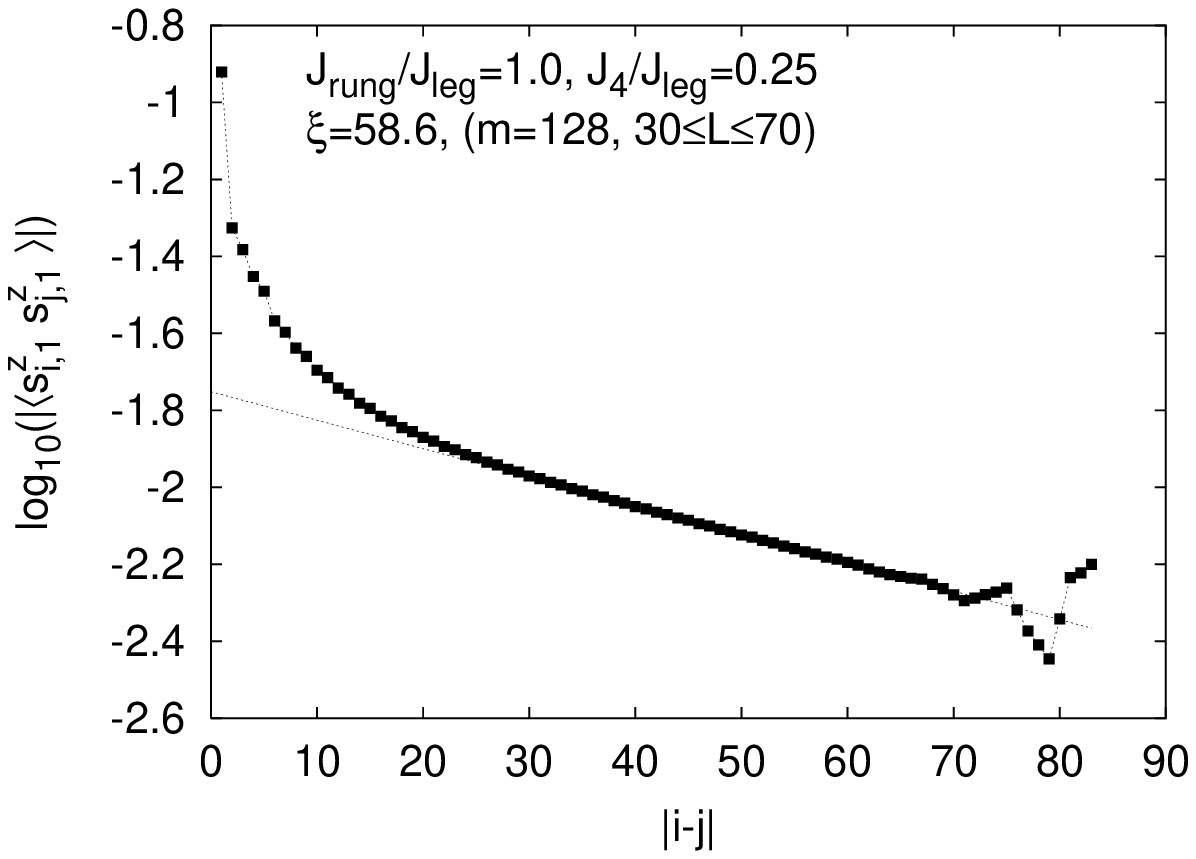}}
  \vspace{0.15\figwl}
  \caption{\label{fig:corr_a1J4025}
     Spin-pair correlation function as a function of distance between 
     $S_{i,1}^z$ and $S_{j,1}^z$ for $J_4/J_{\rm leg}=0.25$ and $a=1$.}
  \end{center}
\end{figure}

We estimate the correlation length of the spin-pair correlation function
$\langle S^z_{i,1} S^z_{j,1} \rangle$.
We choose sites $i$ and $j$ such that they are located symmetrically
with respect to the center of the ladder as shown in Fig.\ref{fig:corr}.
We have obtained the same results for $\langle S^z_{i,2} S^z_{j,2} \rangle$ as 
those for $\langle S^z_{i,1} S^z_{j,1} \rangle$, 
and hence we show the results only for $\langle S^z_{i,1} S^z_{j,1} \rangle$.
In Fig.\ref{fig:corr_a1J4025}, the spin-pair correlation function for 
$J_4/J_{\rm leg}=0.25$ and $a=1$ is shown as a function of distance 
between spins as a typical example.
%
For the gapfull case, we assume the asymptotic behavior of the spin-pair 
correlation function as follows:
\begin{equation}
\langle S^z_{i,1} S^z_{j,1} \rangle \propto e^{-\frac{|i-j|}{\xi}}.
\label{eq:correlation_exp}
\end{equation}
On the other hand, we expect the power-law decay of the
spin-pair correlation function in the gapless case as follows:
\begin{equation}
  \langle S^z_{i,1} S^z_{j,1} \rangle \propto |i-j|^{-\eta},
  \label{eq:correlation_pow}
\end{equation}
where $\eta$ means the critical exponent for the correlation function.
Inset of Fig.\ref{fig:corr_a1J4025} is a semilogarithmic plot of the
absolute value of the spin-pair correlation function.
A deviation from linear shape around $|i-j| \sim 80$ is due to
the effect of the open boundary.
We have used a range $30 \leq L \leq 70$ for estimation of the
correlation length $\xi$.
The correlation length $\xi$ is estimated to be 58.6 for 
$J_4/J_{\rm leg}=0.25$ and $a=1$ by use of the least squares method for
the semilogarithmic plot.
This value of the correlation length is much larger than $\xi \simeq 3.2$ for
the AFHL, namely $J_4/J_{\rm leg}=0$ and $a=1$.
In the fitting by the least squares method, we obtain the error sum of squares 
(ESS) is 0.00029 for $J_4/J_{\rm leg}=0.25$ and $a=1$ by assuming the form 
(\ref{eq:correlation_exp}).
On the other hand, we obtain the ESS is 0.0034 by assuming the 
form (\ref{eq:correlation_pow}) with $\eta=0.814$ as the best fitting.
These results suggest that the spin-pair correlation function decays 
exponentially.

\begin{figure}
  \begin{center}
  \includegraphics{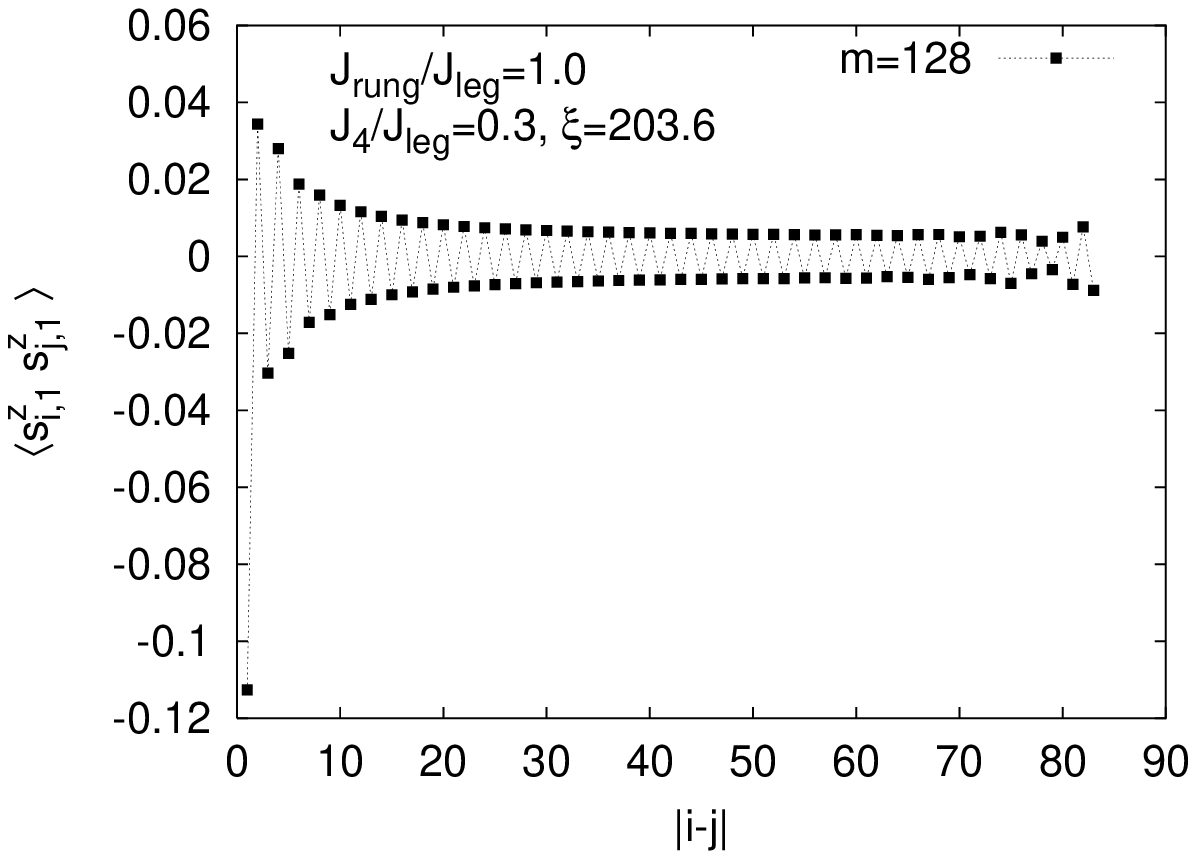}
  \vspace{-0.38\figwl}\\
  \hspace{0.38\figwl}
  \scalebox{0.4}{\includegraphics{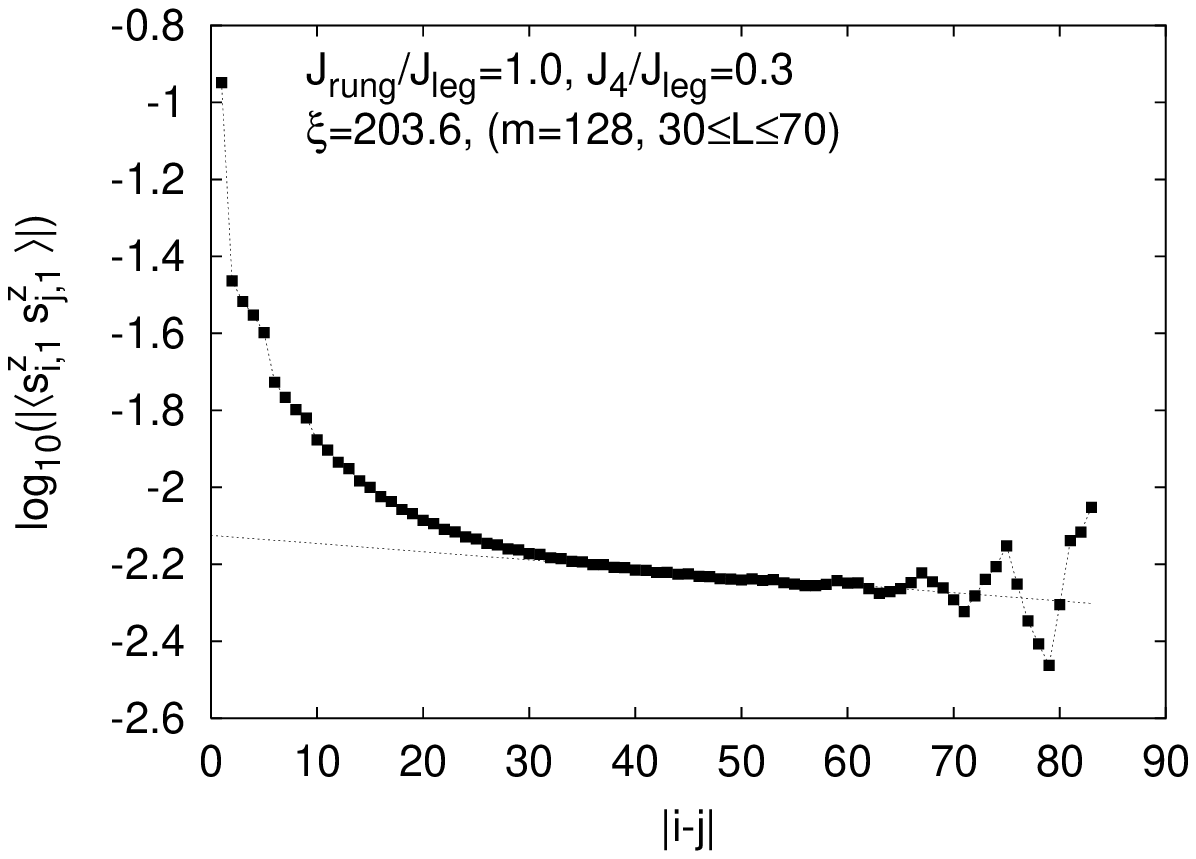}}
  \vspace{0.15\figwl}
  \caption{\label{fig:corr_a1J403}
     Spin-pair correlation function as a function of distance between two spins 
     for $J_4/J_{\rm leg}=0.3$ and $a=1$.}
  \end{center}
\end{figure}

Next, we show the spin-pair correlation function for
$J_4/J_{\rm leg}=0.3$ and $a=1$ in Fig.\ref{fig:corr_a1J403}.
%
We have obtained a quite large value for the correlation length, that is
$\xi=203.6$,by assuming the exponential
decay (\ref{eq:correlation_exp}) where the ESS is 0.0058.
By using the form (\ref{eq:correlation_pow}), we obtain 0.0045 
for the ESS with $\eta=0.242$. 
Then, there is a possibility that the spin-pair correlation function 
decays in the power-law in this case.
These results are consistent with the result that there is no  
spin gap for this value of $J_4/J_{\rm leg}$.
We notice that the value of $\eta \sim \frac{1}{4}$ for this quantum phase transition 
is different from $\eta=1$ for $S=1/2$ spin
chain \cite{Luther75,Hallberg96}.

\subsection{The string correlation function}
\label{subsec:string}
\begin{figure}
  \begin{center}  
  \vspace*{5mm}
  \scalebox{1.5}{\includegraphics{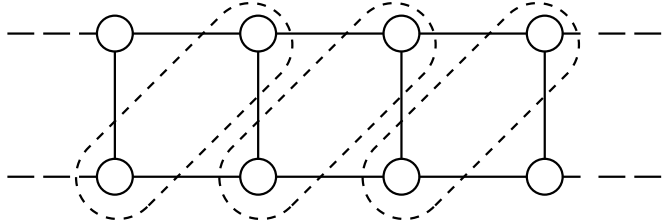}}
  \caption{\label{fig:sto_pair}A single effective $S=1$ spin which consists of
  two $S=1/2$ spins enclosed by dotted line for the antiferromagnetic two-leg spin ladder.}
  \end{center}
\end{figure}

The string correlation function is defined as follows:
\begin{eqnarray}
   g(|i-j|)&=&\left \langle \tilde{S}_i^z \left(\prod_{k=i+1}^{j-1}e^{i \pi \tilde{S}_k^z} \right)
     \tilde{S}^z_{j} \right\rangle \ , \label{eq:sto}
\end{eqnarray}
where
\begin{eqnarray}
   \tilde{S}^z_{i}&=&S^z_{i+1,1}+S^z_{i,2} \  . \label{eq:sto_pair}
\end{eqnarray}
We notice that
$\tilde{S}_i^z$ in eq.(\ref{eq:sto}) consists of two $S=1/2$ spins 
as defined by eq.(\ref{eq:sto_pair}).
These two spins are also illustrated in Fig.\ref{fig:sto_pair}.
This choice of two $S=1/2$ spins was also adopted for the antiferromagnetic 
two-leg ladder without $J_4$ by White \cite{White96},
because this pair of two $S=1/2$ spins is expected to become effectively
a single $S=1$ spin in the antiferromagnetic two-leg ladder.
We choose sites $i$ and $j$ such that they are located symmetrically
with respect to the center of the ladder as shown in Fig.\ref{fig:corr} as well as for
the spin-pair correlation function.

\begin{figure}
\begin{center}
  \includegraphics{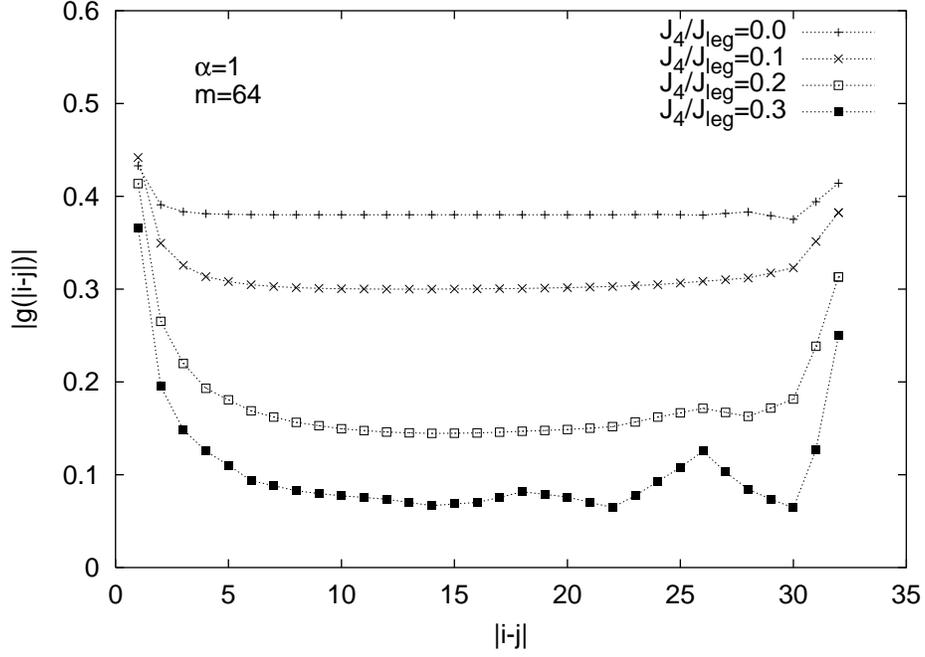}\\
  \hspace{10mm}(a)\\
  \includegraphics{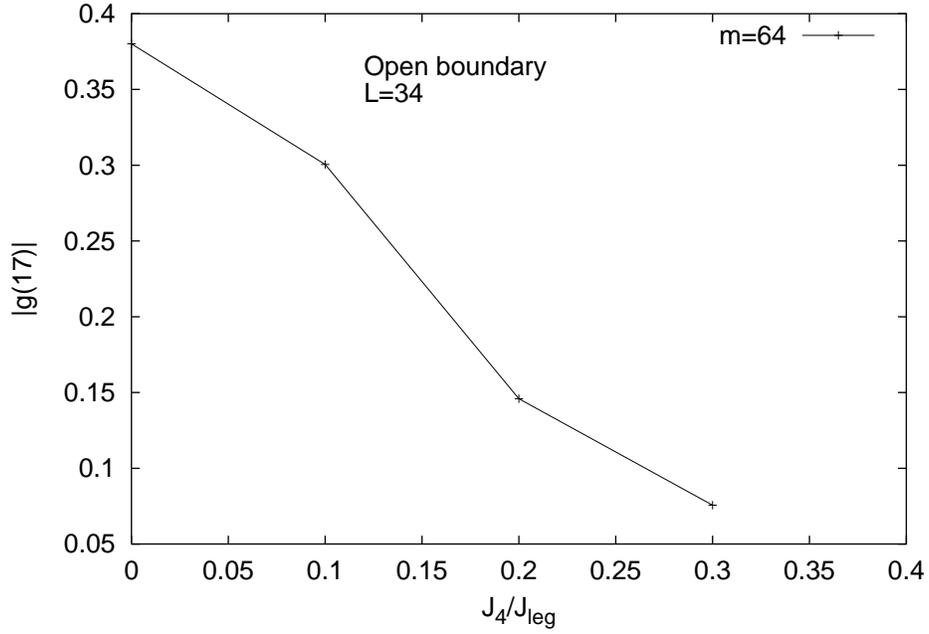}\\
  \vspace*{-3mm}
  \hspace{10mm}(b)
\caption{\label{fig:sto_a1J4dep}
(a) String correlation function for cases in which the value of $J_4/J_{\rm leg}$ is
0.0, 0.1, 0.2 and 0.3 as a function of distance $|i-j|$ between $\tilde{S}_i^z$ and
$\tilde{S}_j^z$. 
(b) Decreasing behavior of the string correlation function for $|i-j|=17$ 
    as a function of $J_4/J_{\rm leg}$.
    Both figures are for the case $a=1$.
}
\end{center}
\end{figure}

In Fig.\ref{fig:sto_a1J4dep}(a), we show values of the string correlation function for
some values of $J_4/J_{\rm leg}$ and $a=1$ as a function of distance $|i-j|$ between effective
spins, $\tilde{S}_i^z$ and $\tilde{S}_j^z$.
For $J_4=0$, namely for the AFHL,
the value of $|g(|i-j|)|$ takes about 0.3801 in range $5 \lesssim |i-j| \lesssim 25$,
where there is no effect of a short distance between $\tilde{S}_i^z$ and $\tilde{S}_j^z$ 
and the open boundary.
This value of $|g(|i-j|)|$ is close to the value of $|g(\infty)|=0.38010765$ 
which was obtained in Ref.\ref{ref:White96}.
The value of $|g(|i-j|)|$ decreases as the value of $J_4$ increases.
We have effect of a short distance due to small values of $|i-j|$ and that of open 
boundary for large values of $|i-j|$.
It should be noticed that these effects increase as the value of $J_4$ increases.
Decreasing behavior of $|g(|i-j|)|$ is shown in Fig.\ref{fig:sto_a1J4dep}(b) as a function
of $J_4/J_{\rm leg}$ by using values of $|g(17)|$ in order to avoid these two effects.
This decreasing behavior suggests that the string correlation function can vanish for 
$J_4/J_{\rm leg} \gtrsim 0.3$.

\begin{figure}
\begin{center}
  \includegraphics{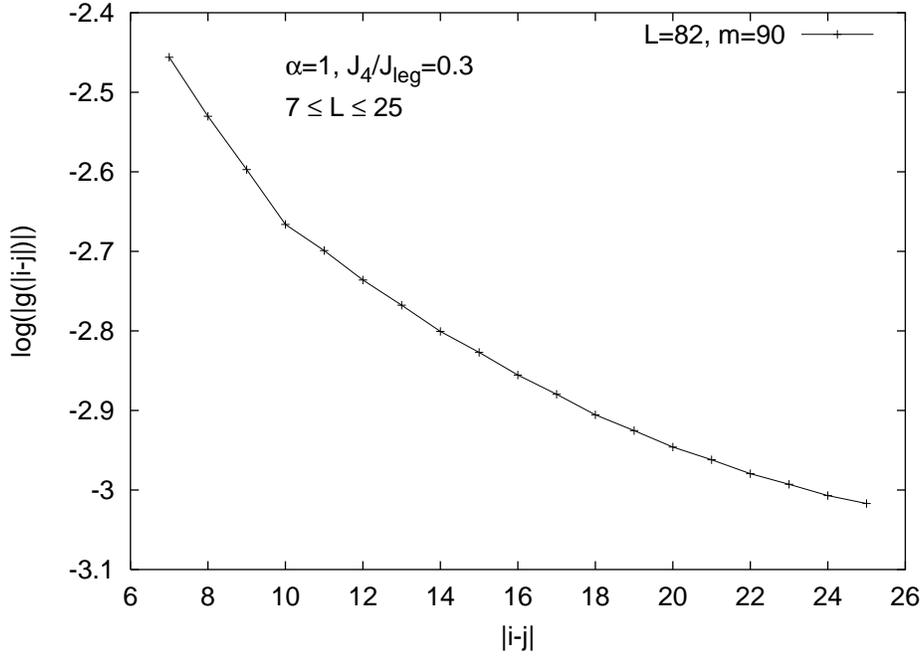}\\
  \hspace{10mm}(a)\\
  \includegraphics{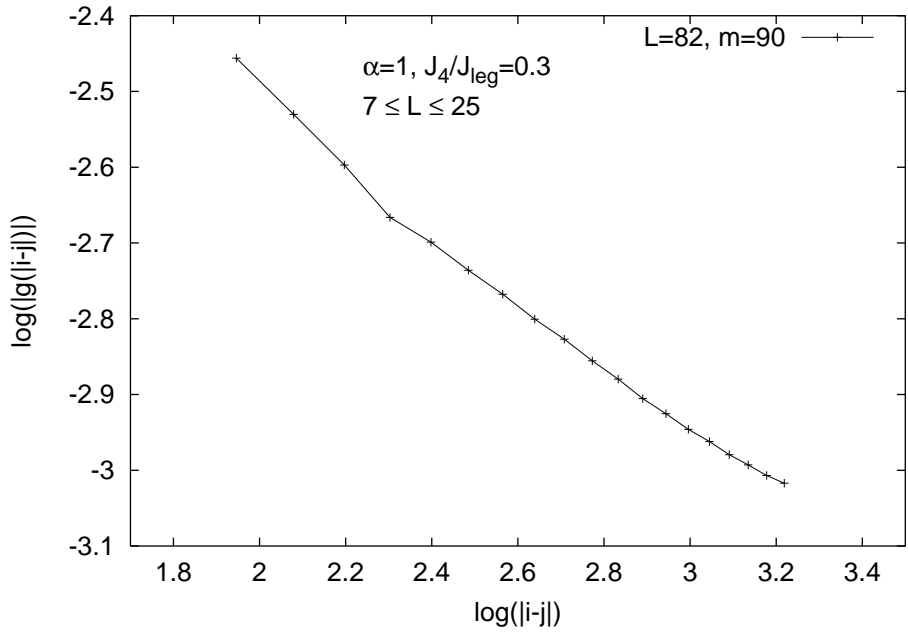}\\
  \vspace{-3mm}
  \hspace{10mm}(b)
\caption{\label{fig:sto_fit}
  (a) The value of $\log\{|g(|i-j|)|\}$ as a function of $|i-j|$.
  (b) The value of $\log\{|g(|i-j|)|\}$ as a function of $\log(|i-j|)$.
}
\end{center}
\end{figure}

We investigate an asymptotic behavior of a decay of the string correlation function
at $a=1$ and $J_4/J_{\rm leg}=0.3$.
We carry out the infinite algorithm of the DMRG method until a system size increases up to $L=82$.
We use a range of system size $7 \leq L \leq 25$ in order to avoid the short distance effect and
the open boundary effect.
In Fig.\ref{fig:sto_fit}(a), values of $\log\{|g(|i-j|)|\}$ are shown as a function of
$|i-j|$.
If the string correlation function decays exponentially as the increasing distance, 
we should have a straight line in this figure.
In Fig.\ref{fig:sto_fit}(b), values of $\log\{|g(|i-j|)|\}$ are shown as a function of
$\log(|i-j|)$.
If the string correlation function decays algebraically as the increasing distance,
we should have a straight line in this figure.
Comparing Fig.\ref{fig:sto_fit}(a) and Fig.\ref{fig:sto_fit}(b), we find that a line in 
Fig.\ref{fig:sto_fit}(b) approximates to a straight line better than that in
Fig.\ref{fig:sto_fit}(a).
Hence, the string correlation function decays algebraically rather than exponentially
at $a=1, J_4/J_{\rm leg}=0.3$.
This result suggests that $a=1, J_4/J_{\rm leg}=0.3$ is a critical point of a quantum phase
transition.

\subsection{Phase diagram}
\label{subsec:phase_diagram}
\begin{figure}
\begin{center}
\resizebox{1.0\figw}{!}{\includegraphics{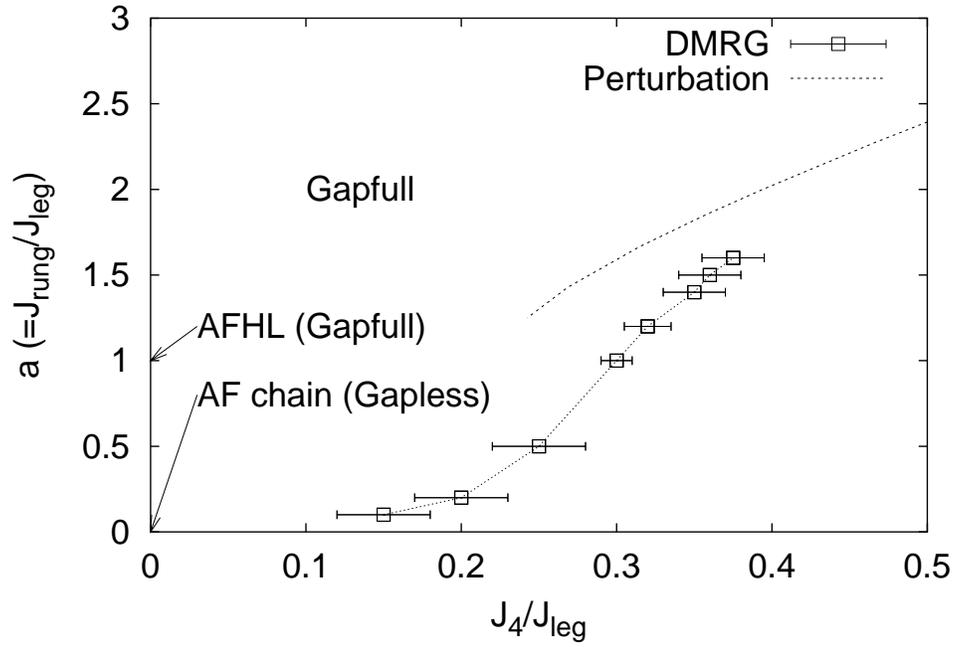}}
\caption{\label{fig:gap_inval-J4}
   A phase diagram in $J_4/J_{\rm leg}-J_{\rm rung}/J_{\rm leg}$ plane.
   A phase boundary is estimated from a value of $J_4$ where the spin gap $\Delta(\infty)$
   vanishes.
   Dotted line is a phase boundary estimated by a perturbation theory \cite{Brehmer99}.
}
\end{center}
\end{figure}

We estimate a phase boundary by searching for values of $J_4/J_{\rm leg}$ at which the
spin gap $\Delta(\infty)$ vanishes for fixed values of $a$.
The obtained phase boundary is shown in Fig.\ref{fig:gap_inval-J4}
in $J_4/J_{\rm leg}-a$ plane,
where  $a=J_{\rm rung}/J_{\rm leg}$.
In this $J_4/J_{\rm leg}-a$ plane, the point (0,1) corresponds to the AFHL which has a
spin gap and a point (0,0) to the antiferromagnetic chain which does not have a spin gap.
As $a$ decreases, the critical value of $J_4$ decreases.
In the present study, we are not able to conclude whether we have a finite range of 
$J_4$ with a finite spin gap at $a=0$ or not.

A dotted line in Fig.\ref{fig:gap_inval-J4} shows a phase boundary 
estimated from the spectrum of the lowest triplet excitation by Brehmer 
{\it et al.} \cite{Brehmer99} who obtained it by a
perturbation theory assuming $J_{\rm leg}/J_{\rm rung}\ll 1$ and 
$J_4/J_{\rm rung} \ll 1$.
In the limit $J_{\rm leg}=0$ and $J_4=0$, the singlet dimers are located every rung of 
the ladder.
Their result was obtained by including terms of third order in
$J_{\rm leg}$ and $J_4$.
Although we can not compare the results by the perturbation theory with our
results by the DMRG method near the region where $J_{\rm rung}/J_{\rm leg} \simeq 1$ and 
$0.25 \lesssim J_4/J_{\rm leg} \lesssim 0.35$, 
both results indicate that the gapfull region broadens as the value of
$J_{\rm rung}/J_{\rm leg}$ increases.

In a region where the value of $J_4/J_{\rm leg}$ is lager than that of the phase boundary, 
the DMRG method becomes unstable.
Hence it remains an open question whether we have a finite spin gap or not in that region.


\section{Conclusion}
\label{sec:conclusion}
We have investigated 
a quantum phase transition for
the antiferromagnetic Heisenberg ladder 
with the cyclic four-spin exchange interaction 
by using the DMRG method; its interaction constant is denoted by $J_4$.
The infinite algorithm with open boundary conditions has been used in order to 
calculate the spin gap, the spin-pair correlation function and the string 
correlation function.
For $J_4=0$, we have not found the gapless region by varying the
value of $a=J_{\rm rung}/J_{\rm leg}$.
This result is consistent with the assertion given by Barnes {\it et al.}
\cite{Barnes93}; the AFHL has a spin gap for all nonzero 
interchain interaction.
On the other hand in the case of $a=1$ and $J_4>0$,
we have found that the spin gap vanishes at $J_4/J_{\rm leg} \simeq 0.3$.
At this point, a larger value of correlation length than the system size 
is found by assuming exponential decay of the spin-pair correlation function.
We have obtained a better fitting by assuming a power-law decay of the 
spin-pair correlation function.
We have found that the string correlation function decays algebraically rather
than exponentially at this point.
These results suggest that 
there is the spin gap for $J_4/J_{\rm leg} \lesssim 0.3$ and
we have a quantum phase transition at
$J_4/J_{\rm leg} \simeq 0.3$ when $J_{\rm rung}=J_{\rm leg}$.
This is a contrast to the result that 
the spin gap remains finite even at phase transition points 
for the system with a combination of biquadratic terms,
which are not related to the cyclic four spin
exchange interaction \cite{Kolezhuk98}.

We have estimated the phase boundary for the
$J_4/J_{\rm leg}-a$ plane and found that it is consistent with the 
result by the perturbation theory for
$J_{\rm rung}/J_{\rm leg}>1$ \cite{Brehmer99}.
Although we have obtained $\eta \sim \frac{1}{4}$ at $a=1$ and $J_4/J_{\rm leg}=0.3$,
estimation of critical exponents remains as a future work 
in order to argue the type of the critical behavior.

The authors wish to thank T. Sakai and H. Yokoyama for fruitful 
discussions. 
This work was supported by the Grant-in-Aid for Science Research 
from the Ministry of Education, Science and Culture(11780183).

\clearpage

\end{document}